\documentclass[useAMS, usegraphicx, usenatbib]{mn2e}
\usepackage{aas_macros}
\usepackage{amsmath}
\usepackage{graphicx}
\usepackage{natbib}
\newcommand{\kunit}{\,h\,\mathrm{Mpc}^{-1}}
\newcommand{\runit}{\,h^{-1}\,\mathrm{Mpc}}
\newcommand{\runitk}{\,h^{-1}\,\mathrm{kpc}}
\newcommand{\munit}{\,h^{-1}\,\mathrm{M}_{\sun}}

\setcitestyle{authoryear,round,comma,aysep={},yysep={,},notesep={}}
\title[Halo matter and the power spectrum]{The contributions of matter inside and outside of haloes to the matter power spectrum}
\author[M. P. van Daalen \& J. Schaye]{Marcel P. van Daalen$^{1,2,3}$\thanks{E-mail: marcel@berkeley.edu} and Joop Schaye$^{1}$\\\\
$^1$Leiden Observatory, Leiden University, P.O. Box 9513, 2300 RA Leiden, The Netherlands\\
$^2$Max Planck Institute for Astrophysics, Karl-Schwarzschild Stra\ss{}e 1, 85741 Garching, Germany\\
$^3$Department of Astronomy, Theoretical Astrophysics Center, and Lawrence Berkeley National Laboratory,\\
\phantom{$^4$}University of California, Berkeley, CA 94720, USA}
\begin{document}
\pagerange{\pageref{firstpage}--\pageref{lastpage}} \pubyear{2015}
\maketitle
\label{firstpage}
\begin{abstract}
Halo-based models have been successful in predicting the clustering of matter. However, the validity of the postulate that the clustering is fully determined by matter inside haloes remains largely untested, and it is not clear a priori whether non-virialised matter might contribute significantly to the non-linear clustering signal. Here, we investigate the contribution of haloes to the matter power spectrum as a function of both scale and halo mass by combining a set of cosmological N-body simulations to calculate the contributions of different spherical overdensity regions, Friends-of-Friends (FoF) groups and matter outside haloes to the power spectrum. We find that matter inside spherical overdensity regions of size $R_\mathrm{200,mean}$ cannot account for all power for $1 \la k \la 100\kunit$, regardless of the minimum halo mass. At most, it accounts for $95\%$ of the power ($k \ga 20\kunit$). For $2 \la k \la 10\kunit$, haloes with mass $M_\mathrm{200,mean} \la 10^{11}\munit$ contribute negligibly to the power spectrum, and our results appear to be converged with decreasing halo mass.
When haloes are taken to be regions of size $R_\mathrm{200,crit}$, the amount of power unaccounted for is larger on all scales.
Accounting also for matter inside FoF groups but outside $R_\mathrm{200,mean}$ increases the contribution of halo matter on most scales probed here by $5-15\%$. Matter inside FoF groups with $M_\mathrm{200,mean}>10^9\munit$ accounts for essentially all power for $3<k<100\kunit$. We therefore expect halo models that ignore the contribution of matter outside $R_\mathrm{200,mean}$ to overestimate the contribution of haloes of any mass to the power on small scales ($k \ga 1\kunit$).
\end{abstract}

\section{Introduction}
The matter power spectrum, a measure of how matter clusters as a function of scale, is a key observable of our Universe. As future weak lensing experiments which aim to measure this quantity with unprecedented accuracy, such as DES\footnote{\texttt{http://www.darkenergysurvey.org/}}, LSST\footnote{\texttt{http://www.lsst.org/lsst}}, Euclid\footnote{\texttt{http://www.euclid-imaging.net/}} and WFIRST\footnote{\texttt{http://wfirst.gsfc.nasa.gov/}}, draw closer, the precision with which the theoretical matter power spectrum is predicted must also increase. Currently, some of the largest uncertainties on fully non-linear scales come from our incomplete understanding of galaxy formation \citep[e.g.][]{vanDaalen2011}, which can cause large unwanted biases in the cosmological parameters derived from observations. We may be able to correct for these biases using independent measurements of, for example, the large-scale gas distribution, and/or by marginalising over these uncertainties using a halo model based approach. However, for the largest of the future surveys more effective and less model-dependent mitigation strategies than currently exist will be needed \citep[e.g.][]{Semboloni2011,Semboloni2013,Zentner2013,Natarajan2014,MohammedSeljak2014,Eifler2014}.

But even assuming that we can somehow account for the effects of galaxy formation on the distribution of matter, significant challenges remain before we are able to predict the matter power spectrum with the sub-percent accuracy needed to fully exploit future measurements \citep{HutererTakada2005,Hearin2012}. These include converging on the ``true'' simulation parameters in N-body codes, although these too can be marginalised over \citep{Smith2014}. However, with each such marginalisation one should expect the constraining power of the observations to be reduced.

Direct simulations are not the only way to obtain theoretical predictions for the matter power spectrum. Other avenues, such as the analytical halo model (e.g.\ \citealt{Seljak2000}, \citealt{PeacockSmith2000}; see \citealt{CooraySheth2002} for a review), exist and are widely used in clustering studies. The halo model is based on the assumption that all matter is partitioned over dark matter haloes, which finds its origin in the model proposed by \citet[][, hereafter PS]{PressSchechter1974}, later extended by \citet{Bond1991}. The PS formalism is based on the ansatz that the fraction of mass in haloes of mass $M(R)$ is related to the fraction of the volume that contains matter fluctuations $\delta_\mathrm{R}>\delta_\mathrm{crit}$, where $\delta$ is the linear density contrast, $R$ is the smoothing scale and $\delta_\mathrm{crit}$ is the critical, linear density contrast for spherical collapse. If the initial field of matter fluctuations is known, then a halo mass function can be derived from this ansatz, which together with a model for the bias $b(M)$ (the clustering strength of a halo of mass $M$ relative to the clustering of matter) and a description of halo density profiles fully determines the clustering of matter.

Much work has been done to improve the predictions of the halo model since its introduction. More accurate mass functions have been derived based on, for example, ellipsoidal collapse \citep{ShethMoTormen2001}, fits to N-body simulations (e.g.\ \citealt{Jenkins2001,Tinker2008,Bhattacharya2011,Angulo2012,Watson2013}; see \citealt{Murray2013} for a comparison of different models) and simulations taking into account the effects of baryons \citep[e.g.][]{Stanek2009,Sawala2013,Martizzi2014,Cui2014,Velliscig2014}. Similarly, much effort has gone into deriving more accurate (scale-dependent) bias functions \citep[e.g.][]{ShethTormen1999,SeljakWarren2004,Smith2007,Reed2009,Pillepich2010,Tinker2010} and concentration-mass relations for halo profiles \citep[e.g.][]{Bullock2001,Neto2007,Duffy2008,Maccio2008,Ludlow2014}. Current halo models may incorporate additional ingredients like triaxiality, substructure, halo exclusion, primordial non-Gaussianity and baryonic effects \citep[e.g.][]{ShethJain2003,SmithWatts2005,Giocoli2010,Smith2011,Gil-Marin2011}, and fitting formulae based on the halo model have also been developed \citep[e.g.][]{Smith2003,Takahashi2012}.

However, the validity of the postulate that the clustering of matter is fully determined by matter in haloes remains relatively untested. Even though matter is known to occupy non-virialised regions such as filaments, their mass may simply be made up of very small haloes itself, although recent results indicate that part of the dark matter accreted onto haloes is genuinely smooth \citep{AnguloWhite2010a,FakhouriMa2010,Genel2010,Wang2011}. Either way, it is not clear a priori whether this non-virialised matter contributes significantly to the non-linear clustering signal.

Here, we examine the contributions of halo and non-halo mass to the matter power spectrum with the use of a set of N-body simulations. This paper is organised as follows. In \S\ref{sec:methods} we describe our simulations and the employed power spectrum estimator. In \S\ref{sec:results} we investigate the contribution to the redshift zero matter power spectrum of haloes that are defined analogous to the typical halo model approach. We start by looking at the fraction of mass that is in haloes as a function of minimum halo mass and compare to analytic results in \S\ref{subsec:massinhaloes}. Next, in \S\ref{subsec:power_radii}, we examine the contributions of matter in regions with lower overdensities and outside of haloes, as a function of Fourier scale. We also examine what changes when we expand the haloes to include all matter associated to Friends-of-Friends (FoF) groups. In \S\ref{subsec:power_mass}, we make predictions for the contribution of halo matter to the power spectrum as a function of both scale and minimum halo mass, which can serve as a test for halo models aimed at reproducing the clustering of dark matter. Finally, we summarise our findings in \S\ref{sec:summary}.

\begin{table}
\caption{The different simulations employed in this paper. From left to right, the columns list their name, box size, particle mass and maximum proper softening length. All simulations were run with only dark matter particles and a WMAP7 cosmology.}
\centering
\begin{tabular}{r r r r r}
\hline
Name & Box size & Particle & $m_\mathrm{dm}$ & $\epsilon_\mathrm{max}$ \\ [0.25ex]
  & $[h^{-1}\,\mathrm{Mpc}]$ & number & $[h^{-1}\,\mathrm{M}_{\sun}]$ & $[h^{-1}\,\mathrm{kpc}]$ \\ [0.5ex]
\hline
\textit{L400} & $400$ & $1024^3$ & $4.50\times 10^9$ & $4.0$ \\ [1ex]
\textit{L200} & $200$ & $1024^3$ & $5.62\times 10^8$ & $2.0$ \\ [1ex]
\textit{L100} & $100$ & $512^3$ & $5.62\times 10^8$ & $2.0$ \\ [1ex]
\textit{L050} & $50$ & $512^3$ & $7.03\times 10^7$ & $1.0$ \\ [1ex]
\textit{L025} & $25$ & $512^3$ & $8.79\times 10^6$ & $0.5$ \\ [1ex]
\hline
\end{tabular}
\label{simtable}
\end{table}

\section{Method}
\label{sec:methods}
\subsection{Simulations}
\label{subsec:simulations}
We base our analysis on a set of dark matter only runs from the OWLS \citep{Schaye2010} and cosmo-OWLS \citep{LeBrun2014} projects. The simulations were run with a modified version of \textsc{gadget iii}, the smoothed-particle hydrodynamics (SPH) code last described in \citet{Springel2005c}. The cosmological parameters are derived from the Wilkinson Microwave Anisotropy Probe (WMAP) 7-year results \citep{Komatsu2011}, and given by \{$\Omega_\mathrm{m}$, $\Omega_\mathrm{b}$, $\Omega_\mathrm{\Lambda}$, $\sigma_8$, $n_\mathrm{s}$, $h$\} = \{$0.272$, $0.0455$, $0.728$, $0.81$, $0.967$, $0.704$\}.

We generate initial conditions assuming the \citet{EisensteinHu1998} transfer function. Prior to imposing the linear input spectrum, the particles are set up in an initially glass-like state, as described in \citet{White1994}. The particles are then evolved to redshift $z=127$ using the \citet{Zel'dovich1970} approximation.

The relevant parameters of the simulations we employ here are listed in Table~\ref{simtable}. The simulation volumes range from $25\runit$ to $400\runit$. The mass resolution improves by a factor of $8$ with each step, corresponding to an improvement of the spatial resolution by a factor of $2$, from the largest down to the smallest volume. The gravitational forces are softened on a comoving scale of $1/25$ of the initial mean inter-particle spacing, $L/N$, but the softening length is limited to a resolution-dependent maximum physical scale which is reached at $z=2.91$. As we will demonstrate, by combining these simulations, we can accurately determine the matter power spectrum from linear scales up to $k\sim 100\kunit$.

\subsection{Power spectrum calculation}
\label{subsec:calculating}
The matter power spectrum is a measure of the amount of structure that has formed on a given Fourier scale $k$, related to a physical scale $\lambda$ through $k=2\pi/\lambda$. It is defined through the Fourier transform of the density contrast, $\hat{\delta}_\mathbf{k}$. We will present our results in terms of the dimensionless power spectrum, defined in the usual way:
\begin{equation}
\Delta^2(k)=\frac{k^3}{2\pi^2}P(k)=\frac{k^3V}{2\pi^2}\left<|\hat{\delta}_\mathbf{k}|^2\right>_\mathrm{k},
\end{equation}
with $V$ the volume of the simulation under consideration. As all particles have the same mass, the shot noise is simply equal to $<\!|\hat{\delta}_\mathbf{k}|^2\!>_\mathrm{k,shot}=1/N_\mathrm{p}$, with $N_\mathrm{p}$ the number of particles in the simulation. All power spectra presented here have had shot noise subtracted to obtain more accurate results on small scales.

We calculate the matter power spectrum using the publicly available \textsc{f90} package \textsc{powmes} \citep{Colombi2009}. The advantages of \textsc{powmes} stem from the use of the Fourier-Taylor transform, which allows analytical control of the biases introduced, and the use of foldings of the particle distribution, which allow the dynamic range to be extended to arbitrarily high wave numbers while keeping the statistical errors bounded. For a full description of these methods we refer to \citet{Colombi2009}. As in \citet{vanDaalen2011}, we set the grid parameter to $G=256$ and use a folding parameter $F=7$ for the two smallest volumes. To calculate the power spectrum down to similar scales for the $200$ and $400\runit$ boxes, we set $F$ equal to $8$ and $9$, respectively. Our results are insensitive to this choice of parameters.

Both box size and resolution effects lead to an underestimation of the power -- at least on scales where a sufficient number of modes is available so that the effects of mode discreteness can be ignored ($k \ga 8\pi/L$) -- while all simulations show excellent agreement on scales where they overlap (see Figure~\ref{fig:power}). In order to cover the dynamic range from $k=0.01\kunit$ to $100\kunit$, we therefore combine the power spectra of different simulations by always taking the largest value of $\Delta^2(k)$ at each $k$. In the case of the full power spectrum, i.e.\ the power spectrum of all matter, we take the combined power spectrum to be the one predicted by linear theory up to $k=0.12\kunit$, where the power starts to become non-linear. While the largest boxes show excellent agreement with the linear power spectrum on these scales, we wish to avoid box size effects as much as possible. For $k>0.12\kunit$ -- or, in the case of power spectra of subsets, for the smallest $k$-value available -- we individually average each power spectrum over each of $25$ bins $k_i$ spaced equally in Fourier space and assign the combined power spectrum the largest $\Delta^2(k_i)$ derived in this manner between all simulations.

We combine the power spectra of selections of particles (e.g.\ all particles that reside in haloes above a certain mass) in a similar way, but without including the linear theory power spectrum.

Finally, we note that we take the contribution of halo matter to the power spectrum to be the auto-correlation of halo matter only (i.e.\ we do not examine the cross terms of halo and non-halo matter).

\subsection{Halo particle selection}
\label{subsec:haloes}
In the halo model approach, haloes are commonly defined through a spherical overdensity criterion, usually relative to the mean density of the Universe. In order to investigate the contribution of such haloes to the matter power spectrum, we define our haloes consistently.

Overdense regions are identified in our simulations using the Friends-of-Friends algorithm (with linking length 0.2 times the mean interparticle distance), combined with a spherical overdensity finder, as implemented in the \textsc{subfind} algorithm \citep{Springel2001}. The centre of each region is taken to be the minimum of the gravitational potential. We define a halo as a spherical region with an internal mass overdensity of $200\times \Omega_\mathrm{m}\rho_\mathrm{crit}$, where $\rho_\mathrm{crit}$ is the critical density of the Universe. These haloes therefore have a mass equal to:
\begin{equation}
M_{200} = M_\mathrm{200,mean} = 200\times\frac{4\pi}{3}\Omega_\mathrm{m}\rho_\mathrm{crit} R_{200}^3,
\label{m200}
\end{equation}
where $R_{200}=R_\mathrm{200,mean}$ is the radius of the region. In the remainder of the paper, we will define halo particles as any particle with a distance $R<R_{200}$ from any halo centre. All other particles are treated as non-halo particles, irrespective of their possible FoF group membership, or having been identified as part of a bound subhalo by \textsc{subfind}.

While we focus on halo matter as defined through $R_\mathrm{200}$, we will also briefly discuss the contribution of halo matter to the power spectrum for other overdensity regions and halo definitions (i.e.\ $R_{500}$, $R_{2500}$, $R_\mathrm{200,crit}$ and Friends-of-Friends) during the course of the paper.

\begin{figure}
\begin{center}
\includegraphics[width=1.0\columnwidth, trim=10mm -14mm 5mm -3mm]{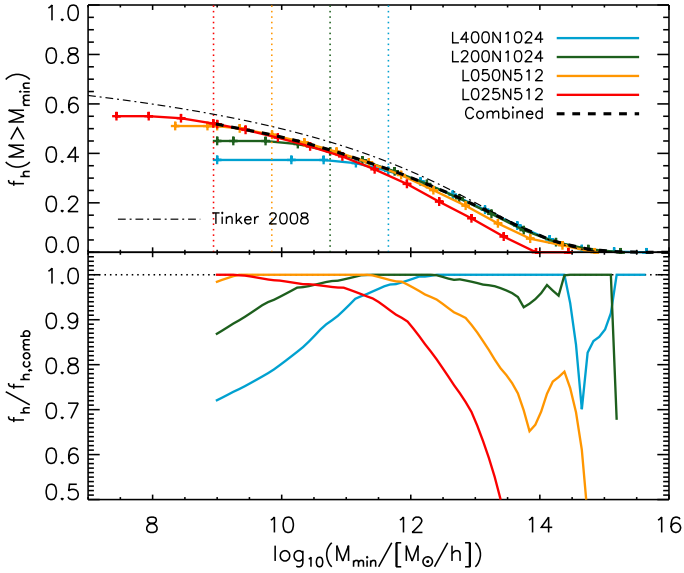}
\caption{\textit{Top panel:} The cumulative fraction of mass inside haloes, $f_\mathrm{h}$, as a function of minimum halo mass, for different collisionless simulations as indicated in the legend. The resolution limit, defined as the mass of haloes containing 100 particles, is shown as a vertical dotted line for each simulation. Below this limit, the fraction of mass in haloes is underestimated. For the two highest-resolution simulations (\textit{L050} and \textit{L025}) these fractions are also significantly underestimated at high masses, as such haloes are under-represented in these small volumes. Between the limits imposed by resolution and box size effects, the simulations are in excellent agreement, and show that the fraction of mass in haloes is $\sim 52\%$ for $M_{200}>10^{9}\munit$. The black dashed line shows the combined result, taking the maximum fraction of mass in haloes between the different simulations at every mass. We also show predictions for the \citet{Tinker2008} mass function as a black dot-dashed line (see main text). \textit{Bottom panel:} The fraction of this combined function, $f_\mathrm{h}/f_\mathrm{h,comb}$, predicted by each simulation.}
\label{fig:massinhaloes}
\end{center}
\end{figure}

\section{Results}
\label{sec:results}
\subsection{Fractional mass in haloes}
\label{subsec:massinhaloes}
We first examine the fraction of the mass that resides in haloes, $f_\mathrm{h}$. As in each simulation there is a lower limit to the masses of haloes that we can reliably resolve, we compute $f_\mathrm{h}$ as a function of the minimum mass of the included haloes. Knowing the minimum resolved masses also allows us to estimate over which halo mass range we can probe the contribution of halo particles to the power spectrum in each simulation.

The results for $f_\mathrm{h}$ are shown in Figure~\ref{fig:massinhaloes}. Different colours are used for each of our four different simulations, as indicated in the legend. Vertical dotted lines denote the masses corresponding to 100 particles. Below this limit the fraction of mass in haloes flattens off, indicating that such low-mass haloes are unresolved. A thick dashed line shows the result of combining the mass fractions of all four simulations for $M_\mathrm{min}>10^9\munit$, through $f_\mathrm{h,comb}=\mathrm{max}(f_{\mathrm{h},i})$, which we consider our best estimate for the true $f_\mathrm{h}$. The bottom panel shows the ratio of $f_\mathrm{h}$ of each simulation to this combined fraction.

At the massive end, the high-resolution but low-volume \textit{L025} and \textit{L050} simulations significantly underestimate $f_\mathrm{h}$. This is most clearly seen in the bottom panel: for \textit{L025} the mass fraction in haloes is significantly underestimated for halo masses $M_{200} \ga 10^{11}\munit$, while for the \textit{L050} box this happens for $M_{200} \ga 10^{12}\munit$. These values correspond to the masses above which the halo mass functions are underestimated for these simulations (not shown). The fluctuations seen in the bottom panel for \textit{L200} where $M_{200}>10^{14}\munit$ are due to the rarity of such massive haloes, but as the fraction of the mass residing in such haloes is $<10\%$ this does not impact our conclusions. All simulations in which haloes at a certain $M_\mathrm{min}$ are both well-resolved and well-represented show excellent agreement for $f_\mathrm{h}(M>M_\mathrm{min})$.

The fraction of mass in haloes increases with decreasing halo mass. Only $\sim 19\%$ of matter is found in groups and clusters ($M_\mathrm{min}>10^{13}\munit$), which increases to $\sim 30\%$ for haloes with masses greater than that of the Milky Way ($M_\mathrm{min}>10^{12}\munit$). But even at the lowest resolved mass of roughly $10^9\munit$, the fraction of mass in haloes is still barely more than $50\%$. We therefore expect a significant contribution from particles in haloes with $M<10^9\munit$ -- and possibly from dark matter particles that do not reside in haloes of any mass -- to the matter power spectrum on large scales, which we calculate in the next section.

For comparison, the top panel of Figure~\ref{fig:massinhaloes} also shows predictions for the fraction of mass in haloes from the \citet{Tinker2008} $M_{200}$ halo mass function. Using the normalized halo mass function fit provided by these authors, we have calculated $f_\mathrm{h}(M>M_\mathrm{min})$ under the standard halo model assumption that all mass resides in haloes (i.e.\ the fit converges to unity when all halo masses are included). The results are shown by the black dot-dashed line. Up to $M_\mathrm{min} \approx 10^{12}\munit$ the relative difference between the \citet{Tinker2008} prediction and our combined result is constant at about $10\%$ before decreasing at higher masses. One possible reason for this discrepancy is that we count matter in regions where haloes overlap only once, whereas double counting is possible when integrating the mass function. However, we have checked that the mass residing in overlap regions in our simulation is always $\la 1.7\%$, with the largest overlap fraction being found for the most massive haloes. The $\la 10\%$ differences found for $f_\mathrm{h}$ are therefore likely due to the non-universality of the halo mass function at this level of precision \citep[e.g.][]{Tinker2008, Murray2013}, or perhaps cosmic variance.

Whether or not $f_\mathrm{h}(M<M_\mathrm{min})$ converges to unity in reality depends on the nature of dark matter, but for perfectly cold dark matter the expectation is that it should. We note, though, that the convergence of $f_\mathrm{h}$ with mass is extremely slow. Taking the \citet{Tinker2008} fit as an example, the fraction of mass in haloes with $M_{200} > 10^9\munit$ is $0.56$, and this number is still only $0.76$ for $M_{200} > 1\munit$. Even for $M_{200} > 1\,\mathrm{kg}$, $f_\mathrm{h}(M<M_\mathrm{min}) \approx 0.88$. The prediction that a significant amount of dark matter is in ultra-small haloes means that the line between halo matter and truly smoothly distributed matter is vague -- but as we will show, it is in at least some cases unnecessary to make the distinction.

Since $f_\mathrm{h}(M>M_\mathrm{min})$ continues to rise down to mass scales that are unresolved by our simulations, we expect to underestimate the total contribution of matter in haloes to the power spectrum. However, as we will see in \S\ref{subsec:power_mass}, this depends on the spatial scale considered. There exists a range in Fourier space where the fraction of power from halo particles converges to values below unity, and the contribution from haloes with masses $M_{200} \la 10^{11}\munit$ is negligible. On scales where this does not hold we can still constrain the contribution from haloes above a certain mass.

In the remainder of the paper, we will only consider particles residing in haloes with $M_{200}>10^9\munit$ to be halo particles, as this corresponds roughly to the smallest haloes we can resolve.

\begin{figure}
\begin{center}
\includegraphics[width=1.0\columnwidth, trim=10mm -13mm 5mm -15mm]{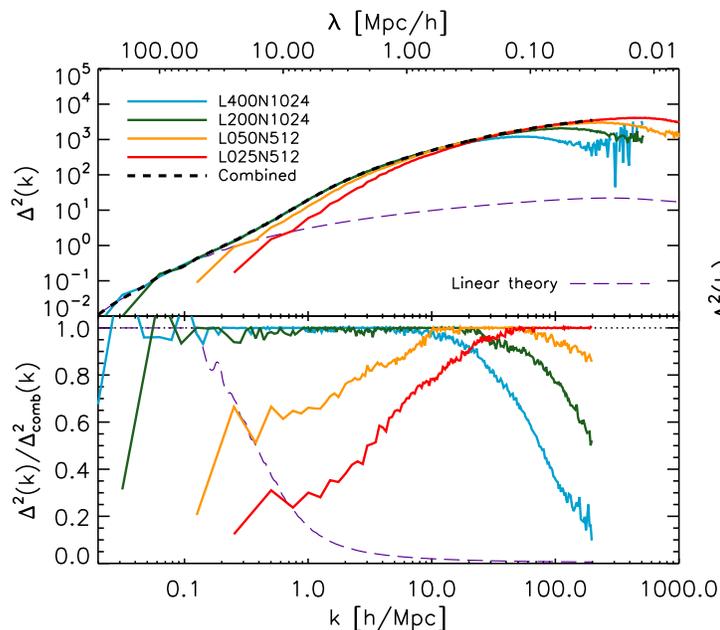}
\caption{\textit{Top panel:} The dimensionless power spectrum derived from each simulation, along with the linear power spectrum (long-dashed purple line) and the combined power spectrum (dashed black line). While the \textit{L025} and \textit{L050} simulations significantly underestimate the power on large scales due to missing modes, their high resolution allows us to accurately extend the power spectrum obtained using the larger volumes up to $k \sim 100\kunit$. The erratic behaviour seen for low-resolution simulations at large $k$ is due to shot noise subtraction. \textit{Bottom panel:} The fraction of power relative to the combined power spectrum for each simulation (as well as for the linear theory prediction). For $k<20\kunit$, multiple simulations show the same result, indicating convergence on these scales.}
\label{fig:power}
\end{center}
\end{figure}

\subsection{Halo contribution to the power spectrum}
\label{subsec:power_radii}
We first show the full dimensionless matter power spectrum, i.e.\ using all particles, in Figure~\ref{fig:power}. Here each simulation is shown using a different colour, and it is immediately clear that no single one is converged over the full range of wavenumbers. The linear theory power spectrum, as generated by the \textsc{f90} package \textsc{camb} \citep[][, version January 2010]{Lewis2000}, is shown as the long dashed purple line. Simulations \textit{L400} and \textit{L200} show good agreement with the linear power spectrum on scales where non-linear evolution is negligible ($k \la 0.12\kunit$) and a sufficient number of modes is available ($k>0.04$ and $0.08\kunit$ respectively, roughly corresponding to $\lambda=0.4\,L$), while \textit{L050} and \textit{L025} show severe box size effects due to their lack of large-scale modes. These box size effects become negligible only for $k>10$ and $k>40\kunit$, respectively.

Due to their finite resolution, all simulations underestimate the power on small scales. Note that shot noise was subtracted from all power spectra shown here, which explains the erratic behaviour of the power spectra on the smallest scales. The underestimation of small-scale power becomes significant already on scales corresponding to $\sim 100$ softening lengths. However, for every wave number $k \la 100\kunit$, there is at least one simulation for which neither box size nor resolution leads to an underestimation of the power at the $\ga 1\%$ level. We therefore combine the different power spectra as described in \S\ref{subsec:calculating} to obtain the combined power spectrum, $P_\mathrm{comb}=\mathrm{max}(P_i)$, shown as the dashed black line.

The bottom panel of Figure~\ref{fig:power} shows the fraction of power predicted by each simulation, as well as the fraction predicted by linear theory, relative to the combined power spectrum. By construction, this fraction is bounded to unity on non-linear scales. Note that on scales $k \la 20\kunit$, the fractions of multiple simulations are within a few percent of unity, indicating convergence on these scales. For smaller scales, however, convergence is uncertain, although based on the results for larger scales we expect our combined power spectrum to be accurate to $\sim 1\%$ up to $k \sim 100\kunit$.

\begin{figure}
\begin{center}
\includegraphics[width=1.0\columnwidth, trim=20mm 8mm 10mm -7mm]{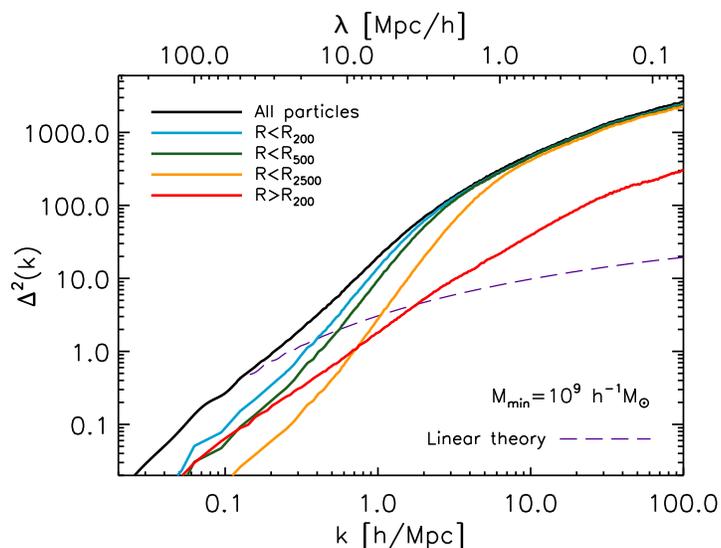}
\caption{The combined power spectrum for different sets of particles: $R<R_{200}$ (halo particles), $R<R_{500}$, $R<R_{2500}$ and $R>R_{200}$ (non-halo particles). Only haloes with $M_\mathrm{200}>M_\mathrm{min}=10^9\munit$ were considered in the cuts made, which in total contain about $52\%$ of all dark matter. The halo particles easily dominate the power on small scales. On linear scales, the contribution of halo and non-halo particles is roughly equal, which is expected as the relative power on these scales depends mostly on the mass in each component. The cross-terms between the halo and non-halo particles (not shown) account for about half the power in the linear and mildly non-linear regimes ($k \la 0.2\kunit$). Note that the horizontal range has been shortened relative to Figure~\ref{fig:power}.}
\label{fig:halopower}
\end{center}
\end{figure}

Next, we repeat this procedure for halo and non-halo particles. We also consider particles within the $R_{500}$ and $R_{2500}$ overdensity regions, defined analogously to $R_{200}$, which probe the inner parts of haloes. As we cannot reliably resolve haloes with less than about 100 particles in any simulation, we only consider the contribution of haloes with masses $M_\mathrm{200}>M_\mathrm{min}=\mathrm{max}\left[10^9\munit,100\,m_\mathrm{dm}\right]$ here, referring to matter in lower-mass haloes as non-halo particles. The results are shown in Figure~\ref{fig:halopower}. Note that for clarity only the combined power spectra are shown, and that the horizontal range has been shortened with respect to Figure~\ref{fig:power}, only showing the range of scales for which we can reliably determine the power spectrum. Furthermore, we note that the power spectrum of each component (e.g.\ particles within $R_{500}$, particles outside of $R_{200}$) was calculated independent of the total matter and not renormalized, in order to facilitate a direct comparison of the power contained in each as a function of scale.

The contribution from halo particles strongly dominates the power on small scales. The halo contribution is in turn dominated by the very inner regions of haloes, at least on scales smaller than the size of these regions. However, towards larger scales this contribution diminishes, and for $k<0.4\kunit$ less than half of the total power is provided by matter in haloes alone. On large scales the significant fraction of the mass that does not reside in haloes with $M_{200}>10^9\munit$ becomes more important, its contribution to the power spectrum increasing to about $20\%$, almost equalling the contribution of halo matter on linear scales. The remaining $\sim 40\%$ of the total matter power on large scales is therefore contributed by the cross-terms of halo and non-halo matter (not shown here).

Note that on the scales shown here, only \textit{L400} and \textit{L200} contribute to the combined power spectrum of non-halo particles. Nonetheless, as Figure~\ref{fig:power} shows that these two simulations are in excellent agreement for $0.4 \la k \la 10\kunit$ even though the mass resolution is eight times worse for \textit{L400}, we have no reason to believe that this component would change significantly on non-linear scales if lower-mass haloes were resolved. On linear scales, however, the contribution of halo matter is mostly determined by the fraction of mass in haloes, which does depend on the minimum halo mass resolved. We will return to this point in \S\ref{subsec:power_mass}.

\begin{figure}
\begin{center}
\includegraphics[width=1.0\columnwidth, trim=20mm 8mm 10mm -7mm]{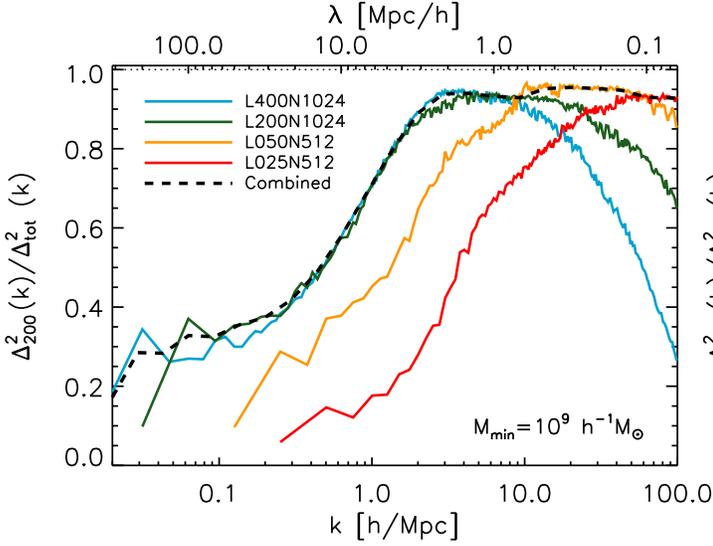}
\caption{The fraction of power within haloes with masses $M_\mathrm{200}>10^9\munit$ as a function of wavenumber. The dashed black line shows the combined power spectrum derived from the smoothed power spectra of the four simulations employed in this paper, each of which is shown as well. The halo contribution rises rapidly down to $\lambda \sim 2\runit$, peaking at $\sim 95\%$ for $k \approx 20\kunit$ ($\lambda \approx 300\runitk$) and remaining roughly constant for larger $k$. On smaller scales the power spectrum is dominated by increasingly smaller haloes, while on the largest scales the contribution of haloes to the power spectrum depends mainly on the total mass fraction in haloes.}
\label{fig:halopower_r200}
\end{center}
\end{figure}

We investigate the contribution of halo matter in more detail in Figure~\ref{fig:halopower_r200}, which shows the ratio of the power spectrum of matter within $R_{200}$ of haloes with masses $M_\mathrm{200}>10^9\munit$ to the power spectrum of all matter. The black dashed line shows the ratio of the combined power spectra, obtained from the smoothed power spectra of all four simulations shown here as described in \S\ref{subsec:calculating}, relative to the combined total power spectrum (black line in Figure~\ref{fig:halopower}). The solid lines show the relative contributions of halo matter separately for each simulation.

The contribution of halo matter to the total power increases with decreasing physical scale. On large (linear) scales, the contribution from haloes seems to converge to $\sim 30\%$, in good agreement with $f_\mathrm{h}(M\!>\!10^9\munit)^2 \approx 0.27$. This is expected, as the contribution of any subset of matter to the power spectrum on sufficiently large scales should scale only with (the square of) the fraction of mass contained in such a subset. However, as the fraction of power in haloes on large scales is fully determined by \textit{L200} and \textit{L400}, with both predicting roughly the same fraction as can be seen in Figure~\ref{fig:halopower_r200}, while the fraction of mass in haloes $M>10^9\munit$ is only accurately measured for \textit{L025}, this correspondence is actually surprising.

On non-linear scales the ratio of the power from halo matter to the total power increases rapidly down to physical scales of $\lambda \sim 2\runit$ ($k \sim 3\kunit$), reaching at most $95\%$, before slowly levelling off towards smaller scales. Note that the combined results are fully determined by \textit{L050} around $k\approx 20\kunit$, where we are unable to show convergence because the resolution of \textit{L200} is too low and the volume of \textit{L025} is too small. However, we will show in \S\ref{subsec:power_mass} that on these scales little would change if higher-resolution simulations were available.

\begin{figure}
\begin{center}
\includegraphics[width=1.0\columnwidth, trim=20mm 8mm 10mm -7mm]{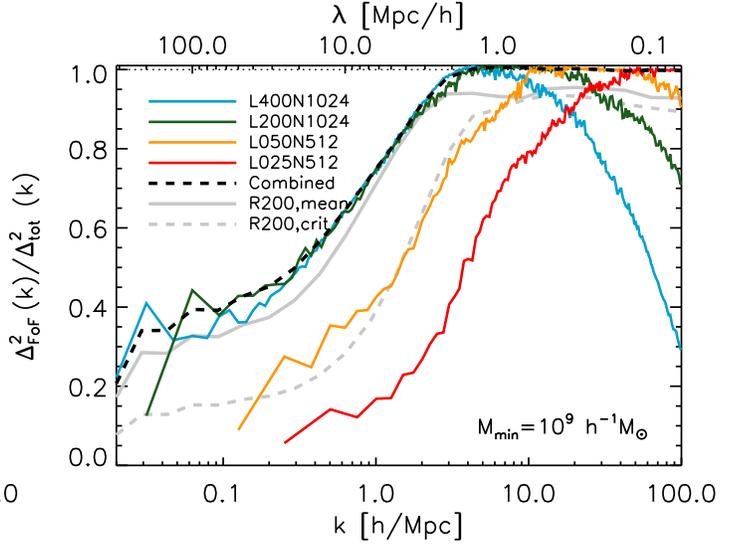}
\caption{As Figure~\ref{fig:halopower_r200}, but now for all mass inside FoF groups with $M_\mathrm{200}>10^9\munit$ (we select on $M_\mathrm{200}$ in order to keep our halo sample identical to the one used in Fig.~\ref{fig:halopower_r200}). While the scale dependence is very similar (i.e.\ a rapid rise down to $\lambda \sim 1\runit$ and roughly constant on smaller scales), the contribution to the power spectrum is higher than for the $R_{200}$ overdensity regions (shown as a solid grey line) on any scale. The contribution of halo matter to the power spectrum is increased by $5-10\%$ on most scales relative to the results of Figure~\ref{fig:halopower_r200}, and matter in FoF groups accounts for essentially all power on scales $k > 3\kunit$. This implies that the $R_{200}$ overdensity regions do not fully capture the halo. The grey dashed line shows the combined result if $R_\mathrm{200,crit}$ is used.}
\label{fig:halopower_fof}
\end{center}
\end{figure}

While \textit{L400} and \textit{L200} are in good agreement for $0.2 \la k \la 10\kunit$, on sub-Mpc scales the contribution of halo matter to the total matter power spectrum starts to show a strong dependence on resolution. On these scales fluctuations within the same halo (i.e.\ the 1-halo term in halo model terminology) dominate the power spectrum, so naturally the contribution to the power will be underestimated on scales $\lambda \la R_\mathrm{200,min}$, where $R_\mathrm{200,min}$ is the virial radius of a halo with the minimum resolved mass, $M_\mathrm{min}$, in that particular simulation. In practice, the power is already significantly underestimated on larger scales. Fortunately, the combination of simulations chosen here still allows us to probe the contribution of halo matter up to $k_\mathrm{max} \sim 100\kunit$.

As the power on sub-Mpc scales is predicted to be dominated by the 1-halo term (which we also observe in our simulations on resolved mass scales), adding lower-mass haloes than those resolved here is expected to have a negligible impact on the measured contribution of halo matter on scales $k_\mathrm{max}<2\pi/R_\mathrm{200,min}$. Therefore, $5-7\%$ of small-scale power is unaccounted for by halo particles, regardless of resolution effects. Note that it is possible that the 2-halo correlations between unresolved and other haloes are responsible for making up the deficit. However, to explain our results these unresolved haloes would have to cluster directly around the resolved haloes, from which a picture arises that is essentially the same as viewing ``smooth'' halo matter as being made up entirely of tiny haloes. We find that it is the cross-term between halo matter and matter \emph{just outside} the $R_{200}$ regions that makes up the deficit.

To demonstrate that this is indeed the case, we calculate the contribution of matter in FoF groups to the total power spectrum, with a mass limit of $M_\mathrm{200}>10^9\munit$. Note that we select on $M_\mathrm{200}$ in order to keep our halo sample identical. The results are shown in Figure~\ref{fig:halopower_fof}. The combined result of Figure~\ref{fig:halopower_r200} is shown as a solid grey line to aid the comparison. While the scale-dependence of the halo contribution for FoF groups is similar to that shown for the $R_{200}$ regions, the halo contribution is significantly larger on all scales (with the exception of scales $k \approx 2\kunit$), and is essentially $100\%$ for $k \ga 3\kunit$. This Fourier scale corresponds to the virial radius of the largest clusters in the simulation.

On scales $k \la 0.3\kunit$, the contribution of matter in FoF groups to the clustering signal is consistently $\sim 20\%$ higher than that of matter in $R_{200}$ haloes. Interestingly, the fraction of mass in FoF groups is only about $4\%$ higher than that in $R_{200}$ haloes (not shown). This implies that the observed increase in the contribution of halo matter to the power spectrum when using the FoF instead of the $R_{200}$ region is not only due to the addition of mass, but mainly due to the addition of clustered material.

Finally, we also show the results if $R_\mathrm{200,crit}$ is used instead (still with $M_\mathrm{200}>10^9\munit$), as a dashed grey line in Figure~\ref{fig:halopower_fof}. As such an overdensity criterion picks out smaller regions than $R_{200}$, containing less mass, the contribution of halo matter to the power spectrum is also smaller, especially on large scales. On sub-Mpc scales, however, the differences are small, with the contribution to the power spectrum of halo matter peaking at $94\%$.

We conclude that what region is chosen to represent a halo has a large impact on the contribution of haloes to the matter power spectrum, in a scale-dependent way. In what follows, we will continue to define haloes using the mean overdensity criterion, as this is typically used in the halo model approach.

\begin{figure*}
\begin{center}
\begin{tabular}{ccc}
\includegraphics[width=0.87\columnwidth, trim=14mm -10mm 51mm -15mm]{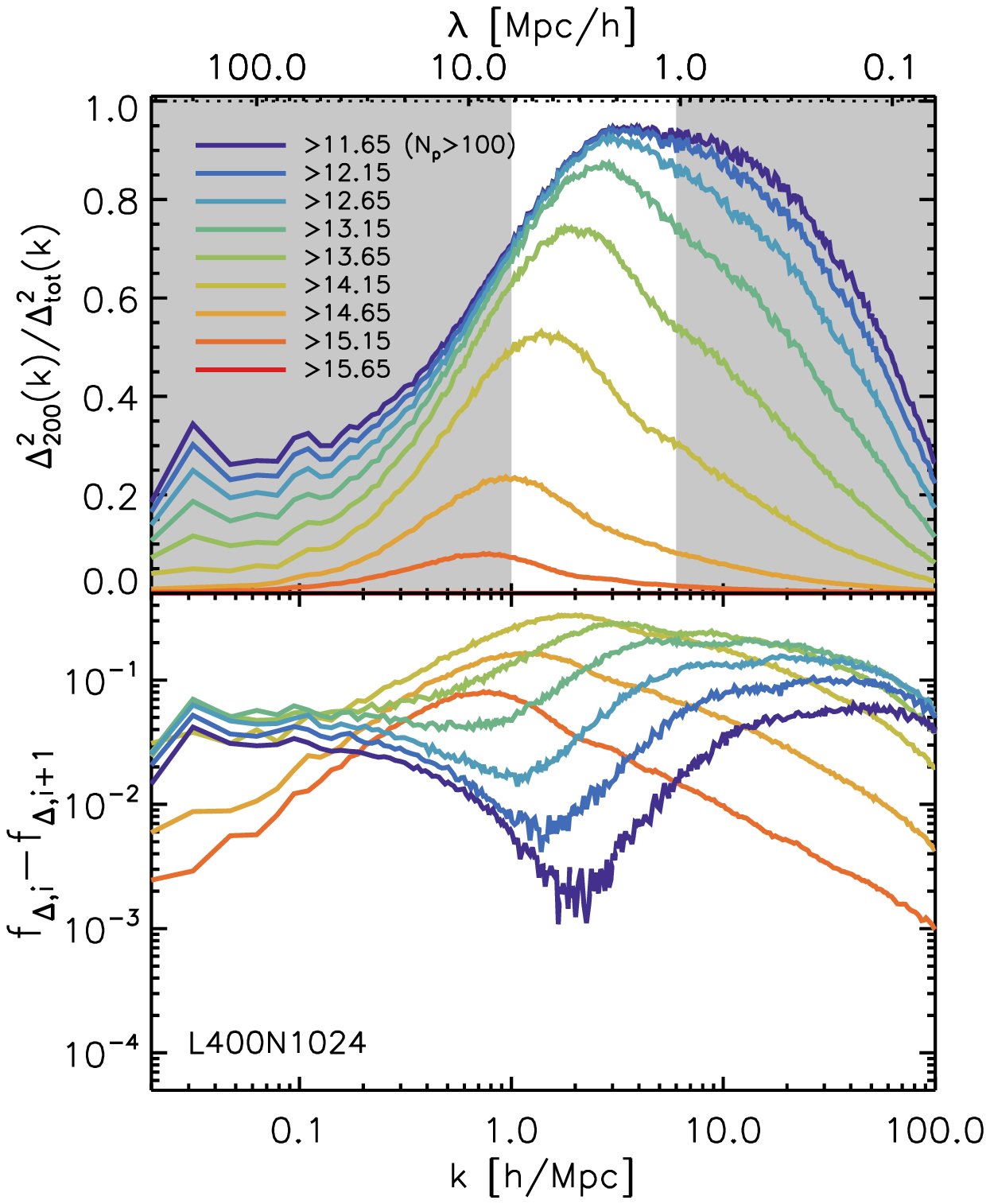} & & \includegraphics[width=0.87\columnwidth, trim=14mm -10mm 51mm -15mm]{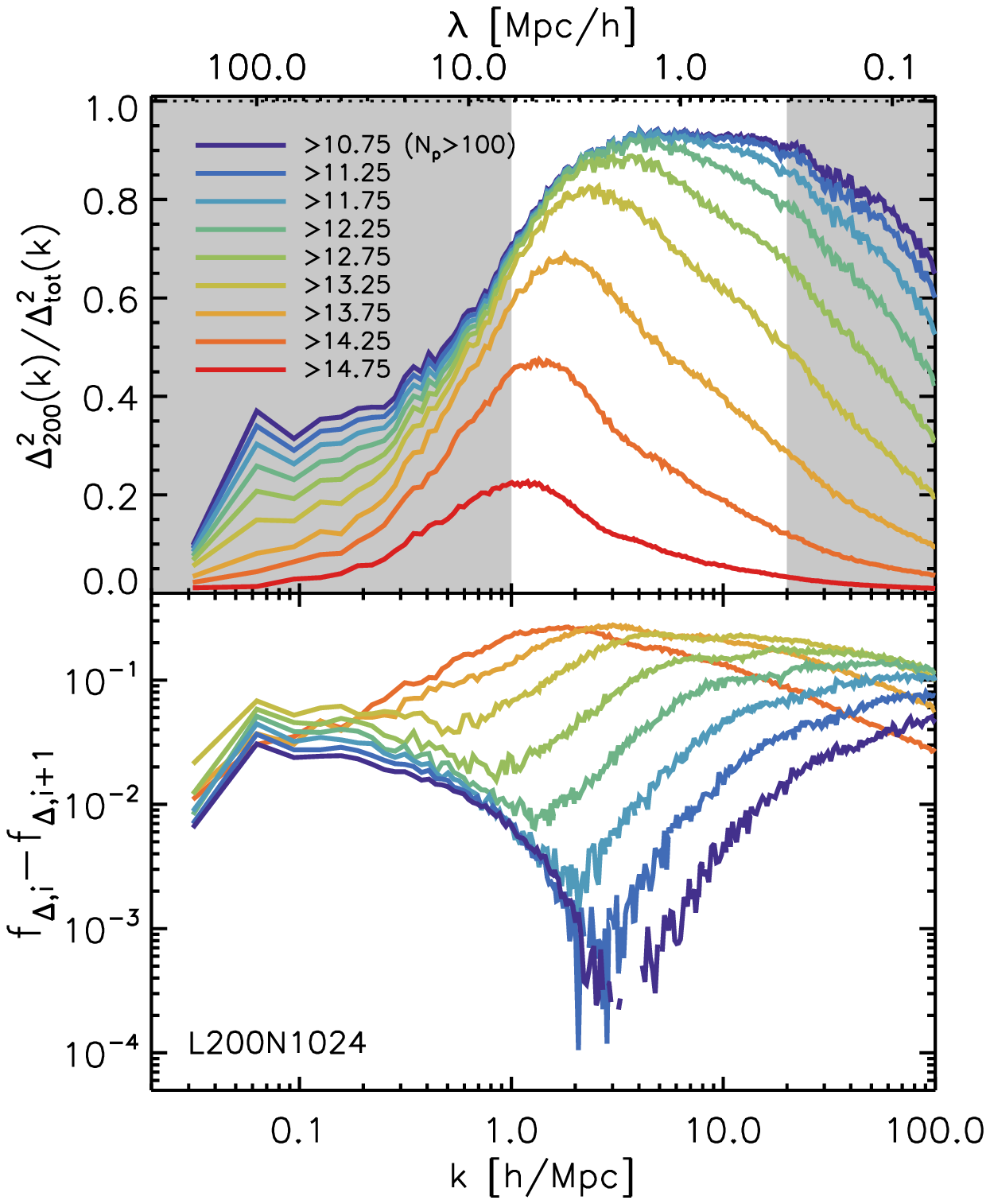}\\
\includegraphics[width=0.87\columnwidth, trim=14mm -10mm 51mm -15mm]{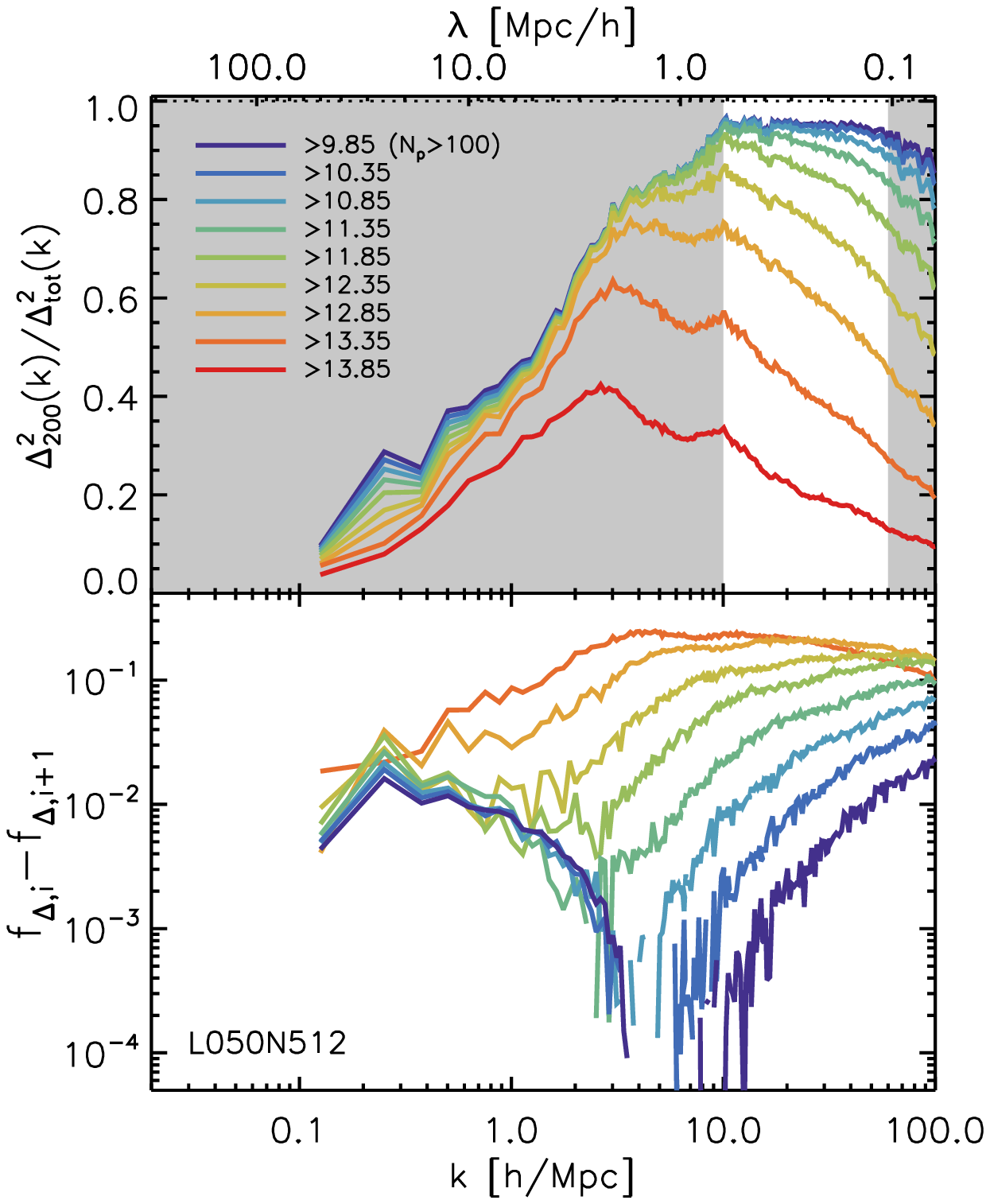} & & \includegraphics[width=0.87\columnwidth, trim=14mm -10mm 51mm -15mm]{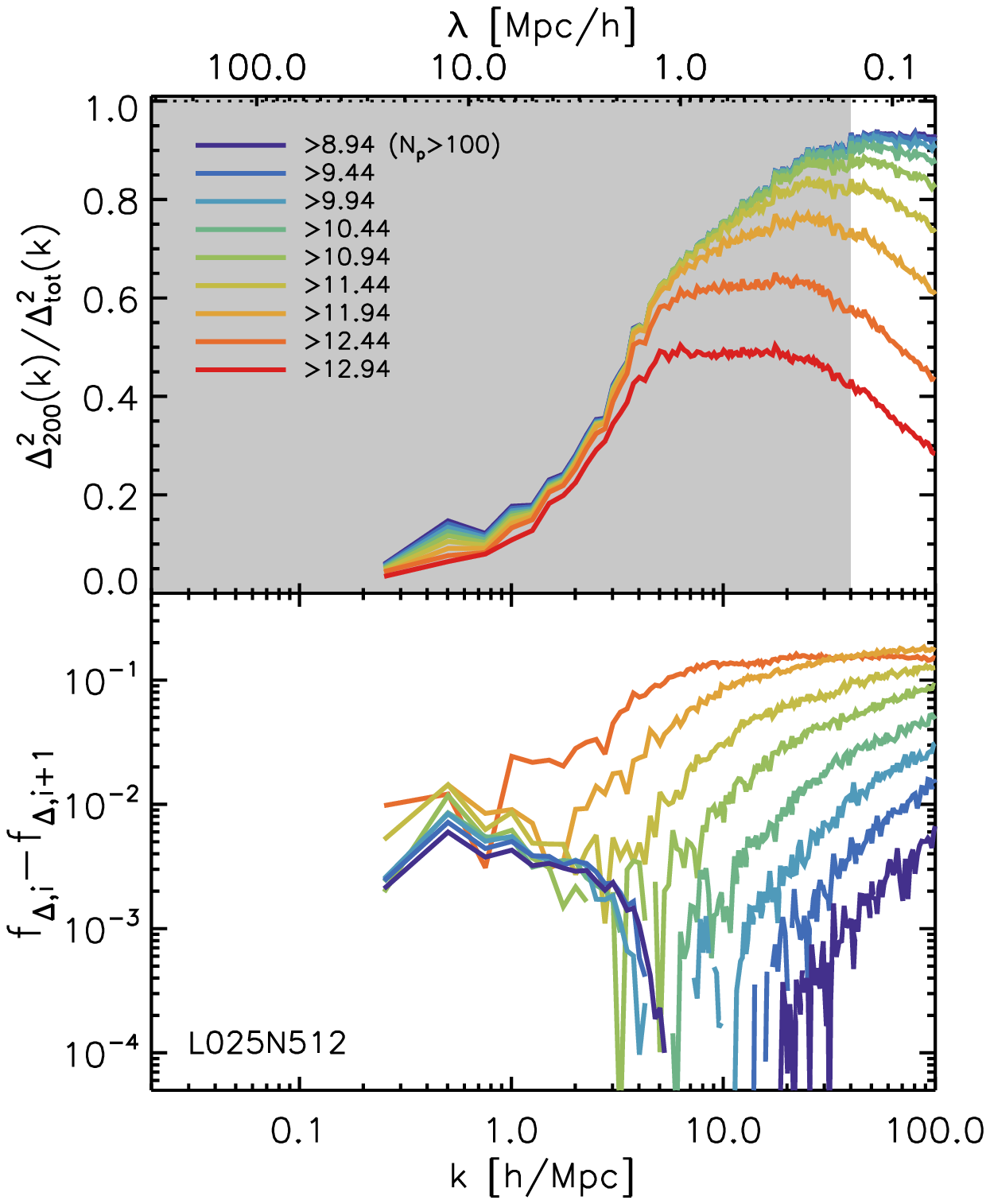}\\
\end{tabular}
\caption{Comparison of the contribution of haloes above a given mass to the matter power spectrum relative to the total combined power spectrum of all simulations, for \textit{L400} (top left), \textit{L200} (top right), \textit{L050} (bottom left) and \textit{L025} (bottom right). The legend shows the minimum halo mass $\log_{10}(M_\mathrm{min}/[\mathrm{M}_{\sun}/h])$. Note that lines of the same colour do not correspond to the same minimum halo mass in the four panels, as the binning is based on the minimum resolved halo mass (see text). The grey regions indicate where box size or resolution effects are $\ga 1\%$ for the full power spectrum; the relative contribution of specific halo masses may be converged on a different range, as can be seen by comparing the panels. The bottom half of each panel shows the difference between consecutive curves in the top panel, i.e.\ the relative contribution added by decreasing the minimum halo mass by half a dex. Several panels show convergence of the fractional power with $M_\mathrm{min}$ to values smaller than unity for wavenumbers that are not greyed out, indicating that the conclusion that matter outside haloes contributes significantly to the power spectrum is robust.}
\label{fig:masspower}
\end{center}
\end{figure*}

\subsubsection{Mass dependence}
\label{subsec:power_mass}
To see which halo masses contribute most to the matter power spectrum as a function of scale, while simultaneously examining the dependence of our results on the mass of the lowest resolved halo, we turn to Figure~\ref{fig:masspower}. Each panel corresponds to a different simulation and each curve to a different minimum halo mass. The halo contributions are shown relative to the combined power spectrum of all matter (black line in Figure~\ref{fig:halopower}).

The legend shows the minimum halo mass, $\log_{10}(M_\mathrm{min}/[\mathrm{M}_{\sun}/h])$, that corresponds to each curve. Note that the minimum masses differ for each simulation, because they start at 100 particles. For each simulation the minimum halo masses of the different curves are half a dex apart.

Grey regions indicate the approximate scales on which the full matter power spectrum of the simulation is not converged to $\sim 1\%$ with respect to the combined one. While this gives an indication of which scales to trust, note that the relative contribution of each halo mass can be converged on a different range of scales.

Finally, the bottom half of each panel shows the difference between consecutive curves, i.e.\ the relative contribution added by decreasing the minimum halo mass by half a dex. Here $f_{\Delta,i} \equiv \Delta_{200,i}^2/\Delta_\mathrm{all}^2$. As we will show shortly, while the relative contributions of haloes of a certain mass shown in the bottom halves of the panels can be compared between different simulations, the same does not hold for the absolute contributions, as box size effects play an important role on a large range of scales.

As Figure~\ref{fig:masspower} shows, all haloes provide a significant contribution to the power on the largest scales ($k \la 0.2\kunit$). It is clear that we cannot claim convergence on these scales. On sub-Mpc scales, low-mass haloes become increasingly important as one moves to larger values of $k$. This means that, as expected, one needs to resolve smaller haloes to obtain convergence on smaller scales. Interestingly, there is a region in between these two regimes where we do see convergence with decreasing halo mass. This is most easily seen in the bottom half of each panel, where the convergence manifests itself as a sharp drop to very low values for low halo masses, indicating that the addition of lower-mass haloes has a negligible effect on the power spectrum.

Focusing on the \textit{L200} simulation (top right panel), we see that it is converged with minimum halo mass on scales $k \sim 4\kunit$, roughly where the contribution from halo matter plateaus. On these scales, haloes with $M_\mathrm{200} \la 10^{11}\munit$ provide a negligible contribution to the power spectrum. This is confirmed by looking at the bottom half of the bottom left panel, showing the results for the \textit{L050} simulation: haloes below $\sim 10^{11}\munit$ do not measurably impact the power for $k \sim 4\kunit$. As we can also see from this panel, the range of scales on which convergence is obtained widens as the minimum resolved halo mass decreases. This convergence, coupled with the result that $\Delta_{200}^2(k)/\Delta_\mathrm{tot}^2(k)<1$, shows that on some scales not all power comes from matter inside $R_{200}$.

\begin{figure}
\begin{center}
\includegraphics[width=1.0\columnwidth, trim=10mm -10mm 51mm -12mm]{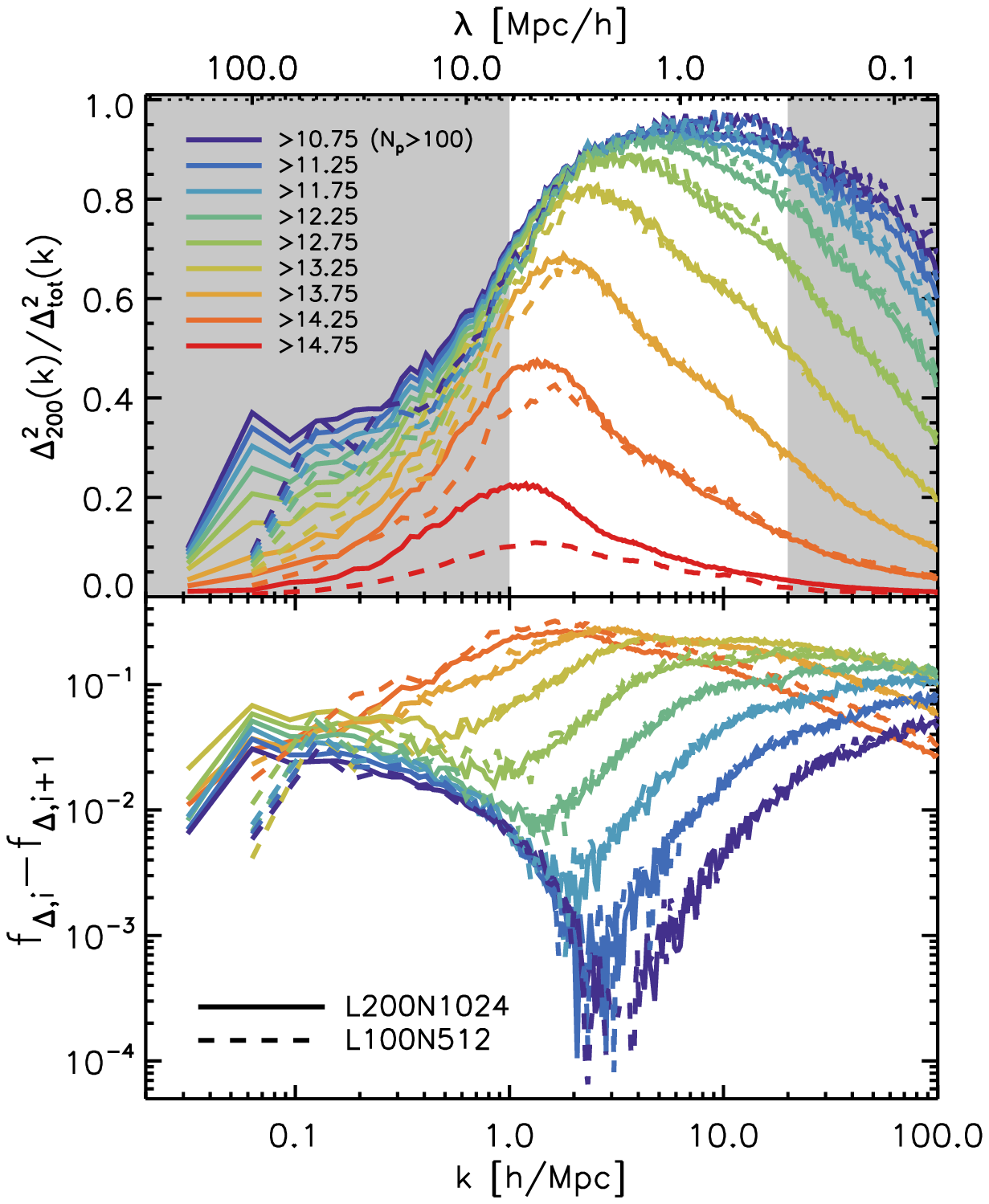}
\caption{As the top-right panel of Figure~\ref{fig:masspower}, but with the results for \textit{L100} added as dashed lines for the same minimum halo masses, showing the effects of box size at fixed resolution. Due to the missing large-scale modes in \textit{L100}, the large-scale contribution is underestimated. Additionally, high-mass haloes are under-represented and the role of low-mass haloes on small scales is overestimated. However, the relative contributions of haloes of a certain mass shown in the bottom half of the figure are in excellent agreement for all but the highest mass bin.}
\label{fig:masspowerboxtest}
\end{center}
\end{figure}

One might worry that the effects of box size skew these results. To test this, we turn to Figure~\ref{fig:masspowerboxtest}, which is identical to the top right panel of Figure~\ref{fig:masspower}, but with the results of \textit{L100} superimposed as dashed lines. The \textit{L100} simulation has $512^3$ particles and therefore the same resolution as \textit{L200}, but in an $8\times$ smaller volume. Comparing the two simulations therefore shows the effects of box size at fixed resolution. On large scales and for high-mass haloes, the contribution of halo matter is underestimated in \textit{L100}, relative to \textit{L200}. This is expected, as large-scale modes are missed in the smaller boxes and massive haloes are under-represented. Meanwhile, the contribution of low-mass haloes on small scales tends to be overestimated, even though the resolution is identical. Interestingly, there are mass and spatial scales where the simulations are in near perfect agreement, such as for a minimum halo mass of $10^{13.75}\munit$ and $k>3\kunit$. Most important, however, is that the contributions from haloes above a given halo mass shown in the bottom half of the panel are in excellent agreement for the two simulations over the entire range of scales, except for the highest mass bins (which are under-represented in \textit{L100}) and the largest modes. This shows that we can still derive the correct contribution of haloes within a certain mass range, and investigate whether we are converged with mass on a certain scale, even when box size effects play a role. For example, while the simulations shown in Figure~\ref{fig:masspower} do not predict exactly the same contribution to the power spectrum for haloes with $M_{200}>10^{11}\munit$, due to the limited box size of \textit{L050} and \textit{L025}, it is clear that the results for the \emph{relative} importance of such haloes are converged. In Appendix~\ref{sec:restests} we investigate the effects of box size and resolution further, and conclude that the contribution of haloes in a certain mass range to the power spectrum is indeed converged with box size over an interestingly large range of scales, while resolution only affects the conclusions on small scales.

Figure~\ref{fig:masspower} makes other interesting predictions as well. For example, comparing again the results for \textit{L200} and \textit{L050}, we see that both simulations agree that most of the power at $k \sim 10\kunit$ comes from haloes with masses $M_{200} \ga 10^{13}\munit$. These group and cluster-scale haloes remain the dominant contributors on somewhat larger scales as well, their contribution peaking around $k = 2-3\kunit$ before gradually falling off. Note that haloes with $M_{200}>10^{13}\munit$ only account for about $19\%$ of the total mass (see Figure~\ref{fig:massinhaloes}). This provides an interesting test for halo models, where the contribution of haloes above a certain mass is strongly dependent on the adopted concentration-mass relation.

\section{Summary \& conclusions}
\label{sec:summary}
In this work we investigated the contribution of haloes to the matter power spectrum as a function of both scale and halo mass, thus testing the assumption typically made by halo-based models that all matter resides in spherical haloes. To do so, we combined a set of cosmological N-body simulations to calculate the contributions of different spherical overdensity regions, FoF groups and matter outside haloes to the power spectrum, paying careful attention to the convergence with both numerical resolution and the size of the simulation volume. As convergence with mass is generally very slow, any claims about the role of haloes or the mass contained in them need to be quoted together with a minimum halo mass in order to have a meaningful interpretation. We note that when we refer to the contribution of halo matter to the matter power spectrum, we consider only the auto-correlation of halo matter, ignoring the cross terms of halo and non-halo matter.

Our findings can be summarised as follows:

\begin{itemize}
\item On scales $k<1\kunit$, haloes -- defined as spherical regions with an enclosed overdensity of $200$ times the mean matter density in the Universe -- with masses $M_{200} \la 10^{9.5}\munit$, which are not resolved here, may contribute significantly to the matter power spectrum. For $2 < k < 60\kunit$, our simulations suggest their contribution to be $<1\%$. A range of scales around $k\sim 4\kunit$ exists for which the contribution of halo matter to the power spectrum appears to be fully converged with decreasing halo mass.

\item Matter within $R_{200}$ alone cannot account for all power for $1 \la k \la 100\kunit$. Its relative contribution increases with increasing Fourier scale, peaking at $\sim 95\%$ around $k=20\kunit$. On smaller scales, its contribution remains roughly constant.

\item When $R_\mathrm{200,crit}$ is used to define haloes instead of the fiducial $R_{200}$, the contribution of haloes to the power spectrum decreases significantly on all scales.

\item Matter just outside the $R_{200}$, but identified as part of FoF groups, provides an important contribution to the power spectrum. Taken together, matter in FoF groups with $M_\mathrm{FoF}>10^9\munit$ accounts for essentially all power for $3 < k < 100\kunit$. Switching from $R_{200}$ to FoF haloes increases the contribution of halo matter on any scale probed here by $5-15\%$.

\item For $2 \la k \la 10\kunit$, haloes below $\sim 10^{11}\munit$ provide a negligible contribution to the power spectrum. The dominant contribution on these scales is provided by haloes with masses $M_{200}\ga 10^{13.5}\munit$, even though such haloes account for only $\sim 13\%$ of the total mass.
\end{itemize}

As we have demonstrated, the halo model assumption that all matter resides in (spherical overdensity) haloes may have significant consequences for the predictions of the matter power spectrum. Specifically, we expect such an approach to overestimate the contribution of haloes to the power on small scales ($k \ga 1\kunit$), mainly because it ignores the contribution of matter just outside $R_{200}$ to the power spectrum.\footnote{The importance to clustering of halo matter outside the virial radius was recently also noted by \citet[][, see their Figure~2]{MohammedSeljak2014}, who found they had to extend the halo profile in their halo model approach to $2\,R_\mathrm{vir}$ in order to better reproduce results from simulations.} While defining haloes to be larger regions similar to FoF groups mitigates the small-scale power deficits, the fact that such regions are often non-virialised and typically non-spherical may lead to other problems.

Clearly, the validity of the postulate that the clustering of matter is fully determined by matter in haloes is strongly dependent on the definition of a halo used -- but it is hard to say what the ``best'' definition to use in this context is. For example, while haloes defined through $R_\mathrm{200,crit}$ will be more compact and therefore have a smaller overlap fraction than $R_\mathrm{200}$ or FoF haloes, their contribution to the power spectrum will be smaller for the same minimum halo mass. And while FoF groups contain nearly all the mass that is important for clustering on small scales, the fact that they are not completely virialised, are non-spherical and have boundaries that do not correspond to a fixed mean overdensity \citep[e.g.][]{More2011} prohibits their use in traditional halo based models.

\section*{Acknowledgements}
The authors thank Simon White for useful discussions, Ian McCarthy for giving us access to and help with the collisionless Cosmo-OWLS simulations, and the OWLS team for running the smaller simulation volumes used here. It is our pleasure to thank the referee, John Peacock, for useful comments that lead to the improvement of this manuscript. The simulations presented here were run on the Cosmology Machine at the Institute for Computational Cosmology in Durham (which is part of the DiRAC Facility jointly funded by STFC, the Large Facilities Capital Fund of BIS, and Durham University) as part of the Virgo Consortium research programme. We gratefully acknowledge support from the European Research Council under the European Union's Seventh Framework Programme (FP7/2007-2013) / ERC Grant agreement 278594-GasAroundGalaxies.
\bibliographystyle{mn2e}
\setlength{\bibhang}{2.0em}
\setlength{\labelwidth}{0.0em}
\bibliography{PhDbib}

\appendix
\section{Additional convergence tests}
\label{sec:restests}
\begin{figure*}
\begin{center}
\begin{tabular}{ccc}
\includegraphics[width=1.0\columnwidth, trim=14mm -10mm 51mm -12mm]{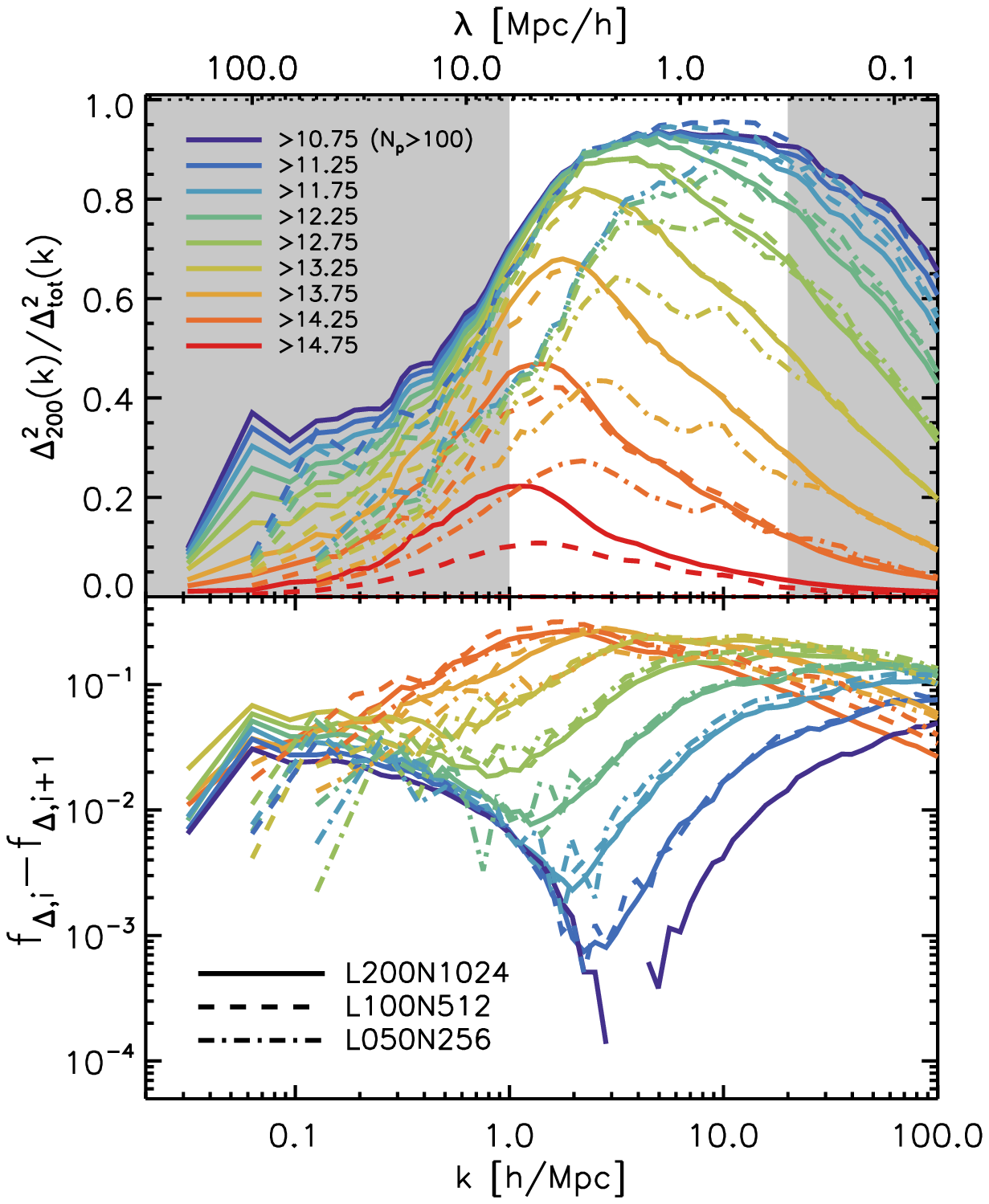} & & \includegraphics[width=1.0\columnwidth, trim=14mm -10mm 51mm -12mm]{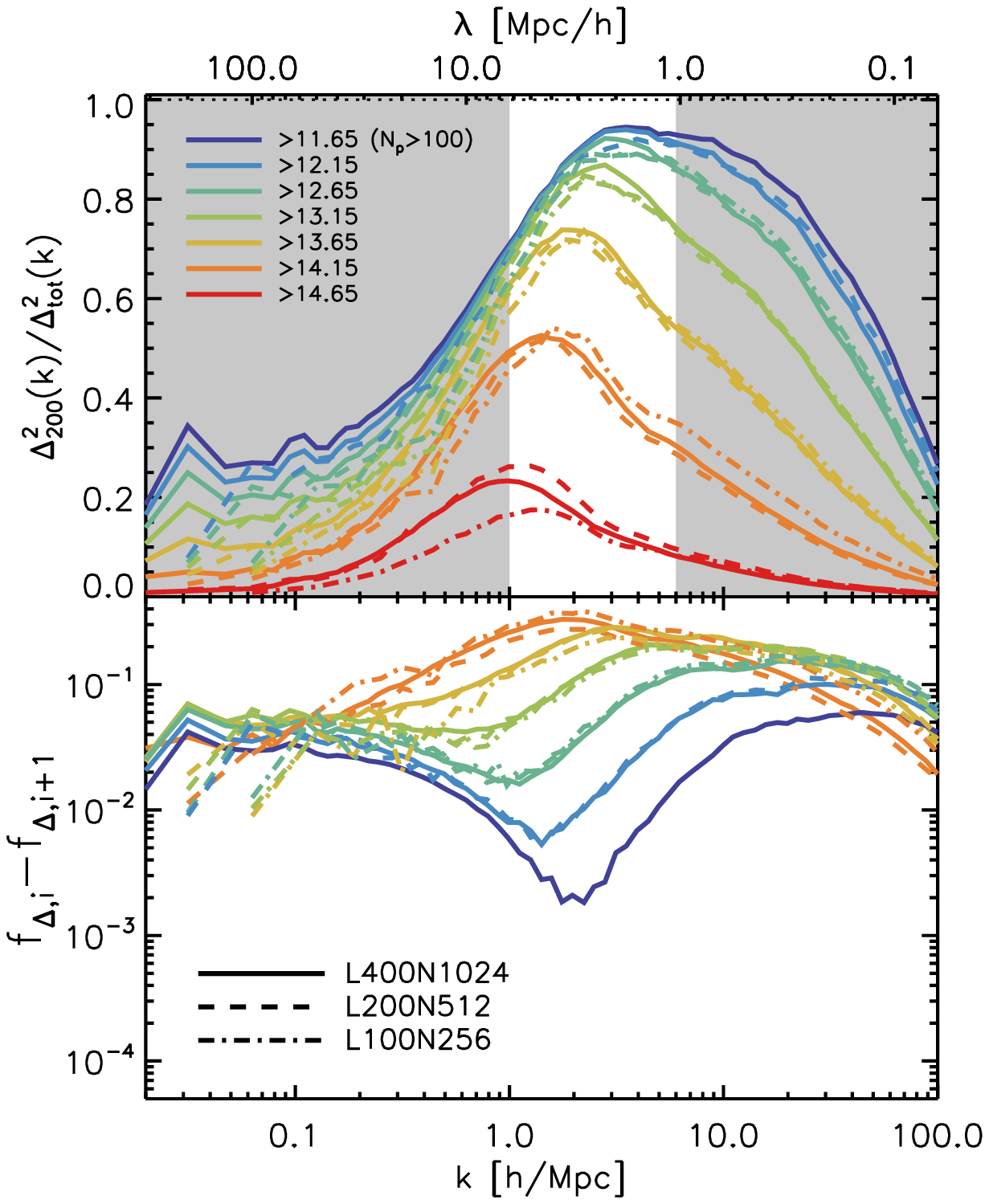}
\end{tabular}
\caption{Convergence test where the box size is changed at fixed resolution, for three high-resolution (\textit{left}) and three low-resolution simulations (\textit{right}). The different simulations are shown as solid (large box), dashed (intermediate box) and dot-dashed (small box) lines. The top panels show the relative contribution of haloes to the power spectrum for different minimum halo masses, shown in the legend as $\log_{10}(M_\mathrm{min}/[\mathrm{M}_{\sun}/h])$. The grey regions indicate where box size or resolution effects are $\ga 1\%$ for the full power spectrum. The curves were smoothed to improve their visibility (see text). The bottom half of each panel shows the difference between consecutive curves in the top panel, i.e.\ the relative contribution added by decreasing the minimum halo mass by half a dex. The effects of decreasing the box size are especially apparent for \textit{L050N256} and for high halo masses, which are under-represented in the smaller boxes. As we saw in Figure~\ref{fig:masspowerboxtest}, lowering the resolution while keeping the box size fixed decreases the relative contribution to the power spectrum for haloes above some mass at large scales, while increasing it at small scales. However, the relative contributions of each \emph{range} in halo mass (shown in the bottom panels) agree very well in each simulation (with the exception of the principal modes and highest mass bin), even when the box size effects are large.}
\label{fig:masspowerboxsize}
\end{center}
\end{figure*}
\begin{figure*}
\begin{center}
\begin{tabular}{ccc}
\includegraphics[width=1.0\columnwidth, trim=14mm -10mm 51mm -12mm]{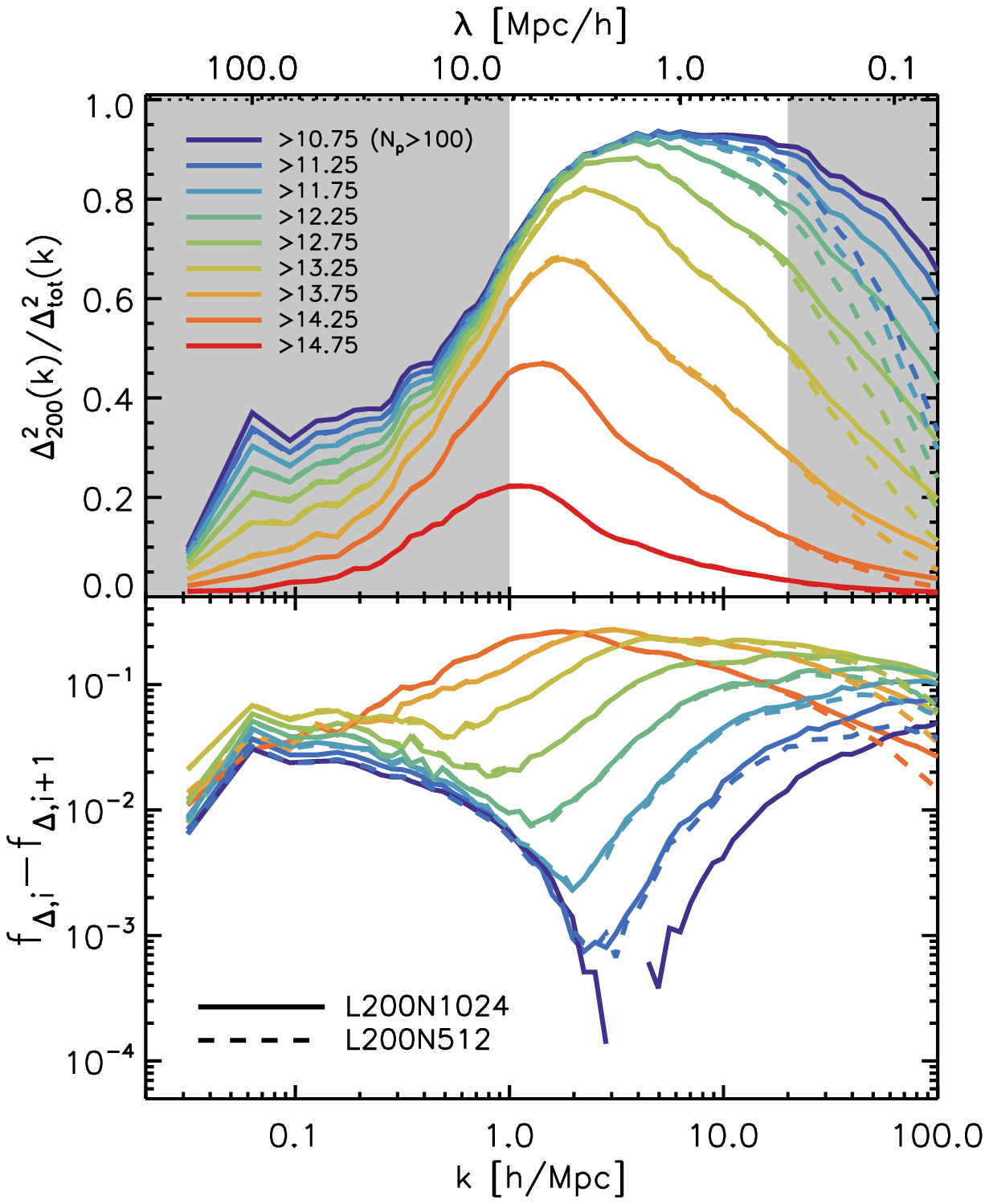} & & \includegraphics[width=1.0\columnwidth, trim=14mm -10mm 51mm -12mm]{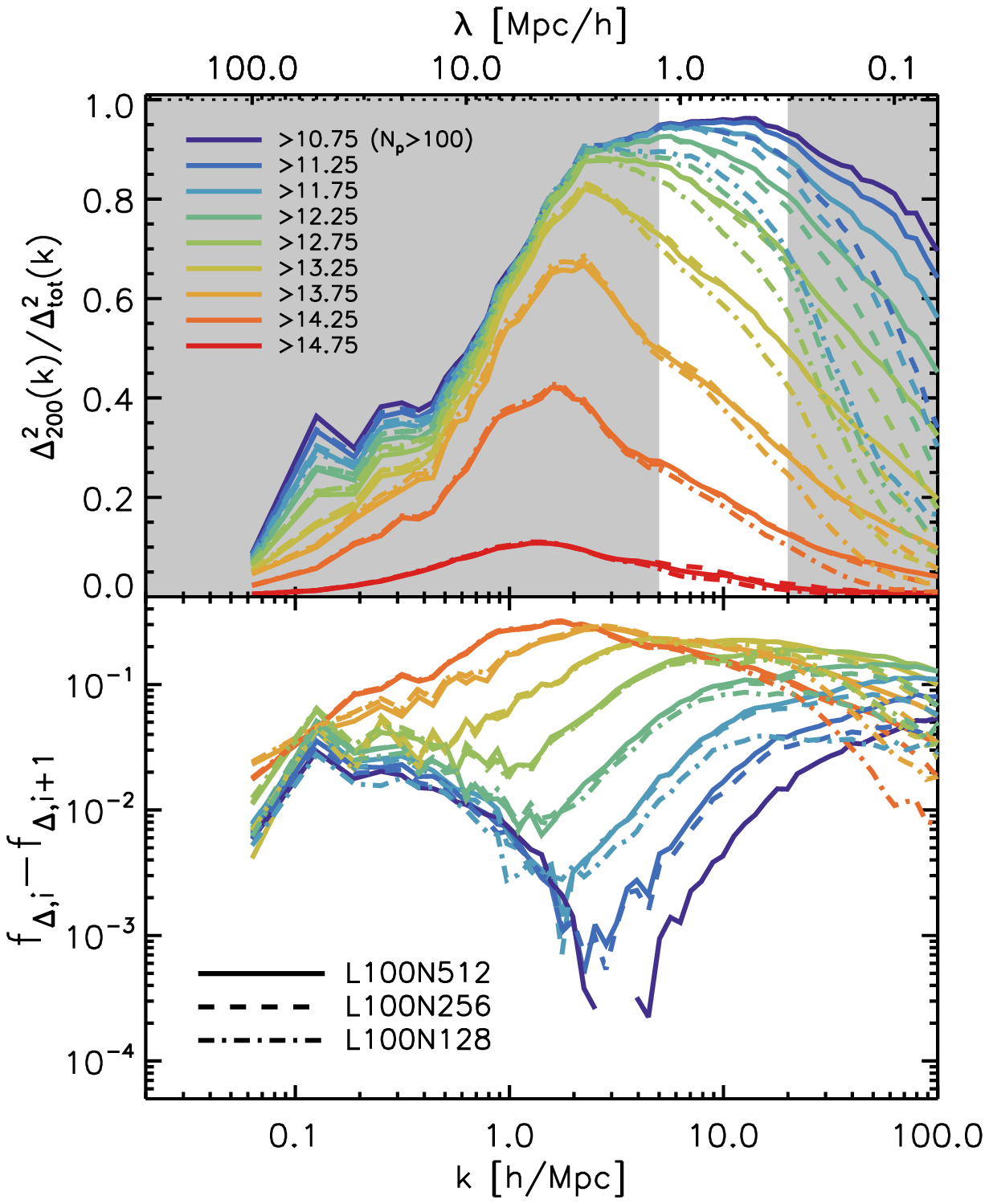}
\end{tabular}
\caption{As Figure~\ref{fig:masspowerboxsize}, but now the resolution is changed at fixed box size, for simulations with box sizes of $200\runit$ (\textit{left}) and $100\runit$ (\textit{right}). Note that we only show lines at masses where haloes are resolved with at least $40$ particles. Lowering the resolution causes the power on small scales and in low-mass haloes to be underestimated, which affects both the relative contribution of haloes above a certain mass (top panels) and in a certain range of halo masses (bottom panels). However, for any halo mass limit a length scale exists down to which the power spectra of simulations with different resolutions show excellent agreement.}
\label{fig:masspowerresolution}
\end{center}
\end{figure*}
Box size and resolution effects can have a large impact on the power spectra measured from simulations. With decreasing box size the number of large-scale modes decreases, leading to an underestimation of the power. Massive haloes also become under-represented, which may lead to an underestimation of the contribution of haloes above a certain mass to the power spectrum. Conversely, with increasing particle mass the minimum mass at which haloes can be reliably resolved increases as well. Here, we investigate these effects independently by changing the box size at fixed resolution and vice versa.

In Figure~\ref{fig:masspowerboxsize} we show the effects of decreasing the box size while keeping the resolution fixed. We do this for two different resolutions: high resolution, based on \textit{L200}, on the left-hand side, and low resolution, based on \textit{L400}, on the right-hand side. For clarity the curves were smoothed by imposing a minimum bin size in $k$ of $0.05\,\mathrm{dex}$ and averaging the power over all modes that fall in each bin. The results for the largest boxes are shown with solid lines, the intermediate-size boxes with dashed lines, and the smallest boxes with dot-dashed lines. The top panels show the relative contributions of haloes above a certain mass to the power spectrum, while the bottom panels show the relative contributions of haloes in mass ranges $0.5\,\mathrm{dex}$ wide.

Looking at the top panels, we see that changing the box size can severely affect the derived contribution of haloes above a given mass. Not only does decreasing the box size lead to a large underestimation of both the power on large scales and of the contribution of massive haloes (especially for \textit{L050N256}), but the power on small scales is simultaneously overestimated in the smaller boxes. However, if we instead turn to the bottom panel we see that the contributions of haloes in a certain mass \emph{range} are much better converged and over almost the entire range, with the exception of the principal modes and the highest mass bin in both figures. The relative contribution of haloes in a certain mass range can therefore be used to investigate convergence, and is the preferred quantity to compare against halo model predictions.

Finally, in Figure~\ref{fig:masspowerresolution} we decrease the resolution while keeping the box size fixed. We show results for simulations with a $200\runit$ box on the left-hand side, and for simulations with a $100\runit$ box on the right-hand side. Both panels show that decreasing the resolution only has a significant effect below some length scale and below some halo mass. For any halo mass limit a length scale exists down to which the power spectra of simulations with different resolutions show excellent agreement.

\bsp
\label{lastpage}
\end{document}